\documentclass[preprint,12pt]{elsarticle}

\usepackage{graphicx}
\usepackage{pdfpages}
\usepackage{subfigure}
\usepackage{float}
\usepackage{lineno}
\usepackage{amsmath}
\usepackage{multirow}
\usepackage{url}

\journal{RDTM}

\biboptions{sort&compress}

\begin{document}
\begin{frontmatter}

		\title{Precise Measurement and Control of Radon Progeny on Detector Surfaces }
		
		\renewcommand{\thefootnote}{\fnsymbol{footnote}}

\author{
		C.B.Z.~Luo$^{a}$,
		C.~Guo$^{b,c,d}\footnote{Corresponding author. E-mail address: guocong@ihep.ac.cn (C.~Guo).}$,
		L.P.~Xiang$^{a}\footnote{Corresponding author. E-mail address: elephantxlp@163.com (L.P.~Xiang).}$,
		Y.H.~Niu$^{e}$,
		F.G.~Mo$^{a}$,
		J.C.~Liu$^{b,c,d}$,
		Y.P.~Zhang$^{b,c,d}$,
		C.G.~Yang$^{b,c,d}$,
		}

\address
{
	${^a}$ School of Nuclear Science and Technology, University of South China, Hengyang, 421001, China\\
	${^b}$ Experimental Physics Division, Institute of High Energy Physics, Chinese Academy of Sciences, Beijing, 100049, China \\
	${^c}$ School of Physics, University of Chinese Academy of Sciences, Beijing, 100049, China \\
	${^d}$ State Key Laboratory of Particle Detection and Electronics, Beijing, 100049, China \\
	${^e}$ Institute of Theoretical Physics, Shanxi University, Taiyuan, 030006, China. \\
}

    \begin{abstract}
    In low-background particle physics experiments, surface deposition of radon progeny presents a significant background challenge. To characterize this contamination, a high-sensitivity surface $\alpha$-activity measurement system was developed, which employs a 3$\times$3 Si-PIN array operating in vacuum to perform $\alpha$-spectroscopy on samples. The system was calibrated using Poly(Methyl MethAcrylate) (PMMA) plates exposed to a controlled high-radon atmosphere, achieving an energy resolution of 2.09 \% for 5.30~MeV $\alpha$ particles and a one-day measurement sensitivity of 1.27~$\mu$Bq/cm$^2$ for $^{210}$Po surface activity. Using this system and a self-built high radon concentration chamber, the deposition behavior of radon progeny on PMMA surfaces was investigated. Results indicate a non-monotonic dependence on exposure time, a significant enhancement of deposition with increasing negative surface electrostatic potential, and a strong modulation by ambient humidity.  This paper details the apparatus design, calibration, and experimental study of radon progeny deposition dynamics on PMMA surfaces.
\end{abstract}


\end{frontmatter}


	\section{Introduction}
	Dark matter research stands as one of the core frontiers in modern high-energy physics. Although cosmological observations provide strong evidence for its existence, the particle nature of dark matter remains unknown. Whether it consists of heavier weakly interacting massive particles, lighter axions, or other candidates, its interaction cross-section with ordinary matter is known to be extremely small. Consequently, direct detection experiments must be conducted in ultra-low background environments.
	
\begin{figure}[htb]
	\centering
	\includegraphics[width=12cm]{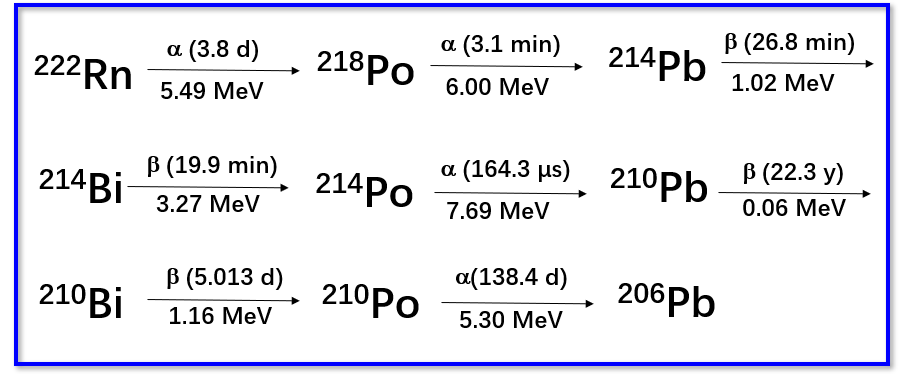}
	\caption{Schematic Diagram of $^{222}$Rn decay chain.}
	\label{decay.png}
	\end{figure}

    In particular, $\alpha$-emitting radionuclides on detector surfaces represent a non-negligible background source~\cite{borexino}. $^{222}$Rn, owing to its relatively high concentration in air and its natural abundance among radon isotopes,  should be strictly controlled. As detailed in Fig.~\ref{decay.png}, $^{222}$Rn has a half-life of 3.82 days and produces $^{210}$Pb in its decay chain, which has a long half-life of 22.3 years. Once $^{222}$Rn or its progeny deposit on detector surfaces, $^{210}$Pb can accumulate within a relatively short time, forming a persistent background that is difficult to eliminate. More critically, $\alpha$ emitters such as $^{210}$Po not only contribute to direct energy deposition but may also generate neutrons via ($\alpha$, n) reactions~\cite{210Po}. These secondary neutrons can mimic nuclear recoil signals in dark matter direct detection experiments~\cite{darkside,pandax,lz} or produce neutron capture signals in liquid scintillator detectors~\cite{juno,sno+}—both of which are challenging to effectively discriminate during later stages of data analysis.

	To address the contamination caused by $^{222}$Rn and its progeny on detector surfaces, international experimental groups have developed corresponding surface treatment techniques tailored to different detector materials. For instance, the GERDA~\cite{zuzel2018removal} experiment employed etching and electropolishing to remove radon progeny from oxygen-free copper and stainless-steel surfaces, while the DEAP-3600~\cite{giampa2018deap} experiment developed a specialized robotic system to ablate the entire inner acrylic surface by approximately 500~$\mu$m. To enhance high-sensitivity measurement and control of surface radon progeny contamination, we have developed a high-sensitivity surface $\alpha$-activity measurement apparatus and studied the influence of different experimental conditions on the deposition activity of radon progeny on Poly(Methyl MethAcrylate) (PMMA) surfaces. The structure of this paper is as follows: Sec.~\ref{sec.2} describes the design of the surface $\alpha$-activity measurement apparatus, Sec.~\ref{sec.3} outlines the calibration process and performance of this system, Sec.~\ref{sec.4} presents experimental results on the deposition activity of radon progeny on PMMA surfaces under various conditions; and Sec.~\ref{sec.5} summarizes the main findings of this research.

	\section{Experimental setup}\label{sec.2}

\begin{figure}[htb]
	\centering
	\includegraphics[width=11cm]{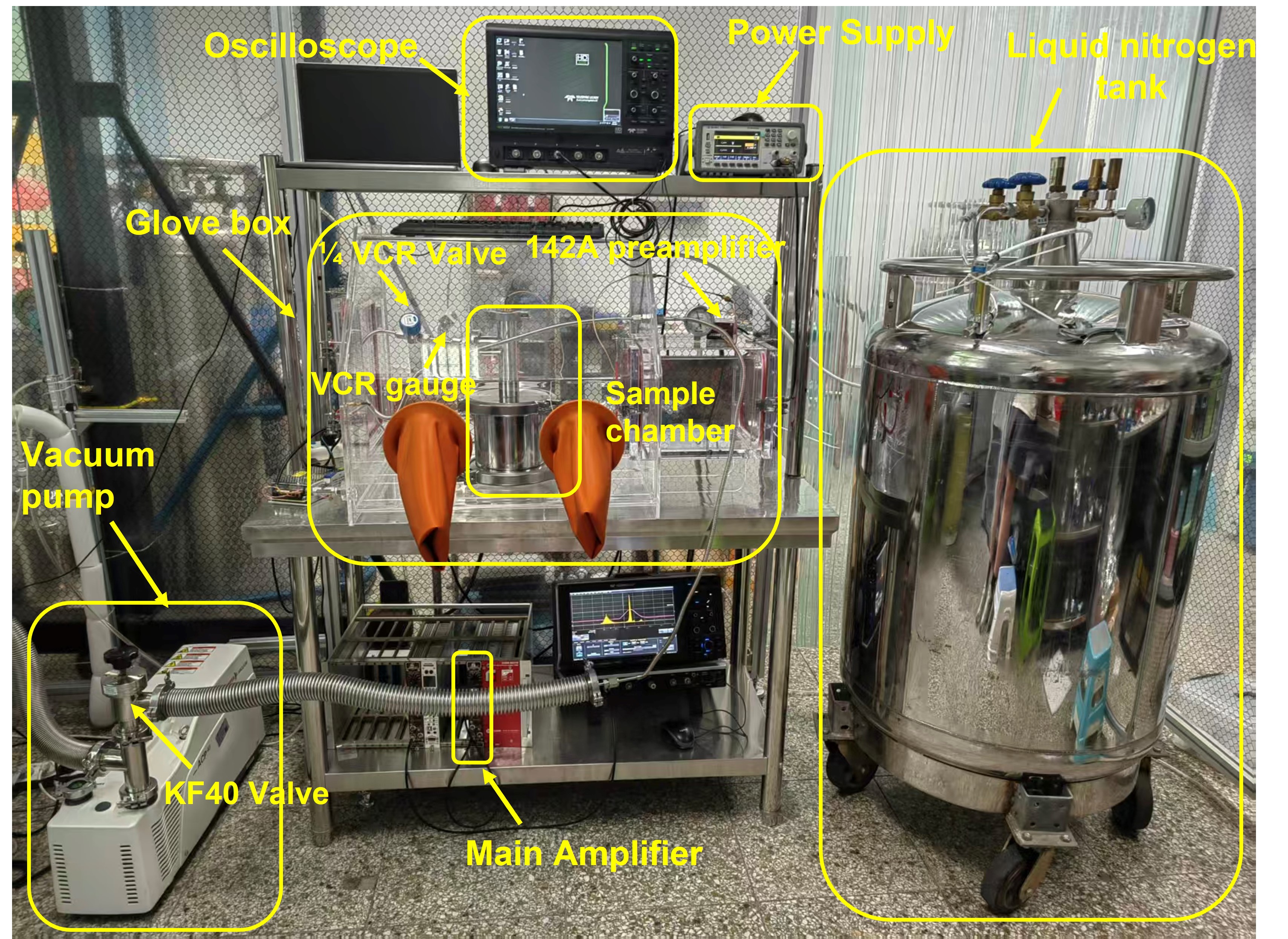}
	\caption{A picture of the Surface $\alpha$ Activity Measurement System.}
	\label{Detection System}
	\end{figure}

In low-background experiments, $^{210}$Po is the principal $\alpha$-emitting nuclide contributing to detector background. Consequently, the primary objective of the instrument developed in this work is to determine the $^{210}$Po activity on material surfaces. $\alpha$ particles emitted from the decay of $^{210}$Po have an energy of 5.30 MeV. Given the short range of $\alpha$ radiation, a Hamamatsu Photonics S3204-09 Si-PIN detector~\cite{citekey} was selected for its detection. To enable accurate energy measurement, the sample must be positioned as close as possible to the detector without risking its contamination. To minimize $\alpha$ particle energy loss, both the Si-PIN detector and the sample are housed in a vacuum chamber. Since air contains substantial amounts of dust and radon gas, prolonged exposure would elevate the detector's background over time. Therefore, the entire measurement system must be operated in a low-background environment. Taking these factors into account, we designed the apparatus illustrated in Fig.~\ref{Detection System}.

\subsection{System Composition}

A schematic of the surface $\alpha$ activity measurement system is shown in Fig.~\ref{Detection System}. It consists of six major components: a sample chamber, a glove box, a vacuum pump, a liquid nitrogen tank, a set of electronics, and an oscilloscope. The details and functions of each component are described below.

  \subsubsection{Sample Chamber}

	\begin{figure}[htb]
		\centering
		\includegraphics[width=10cm]{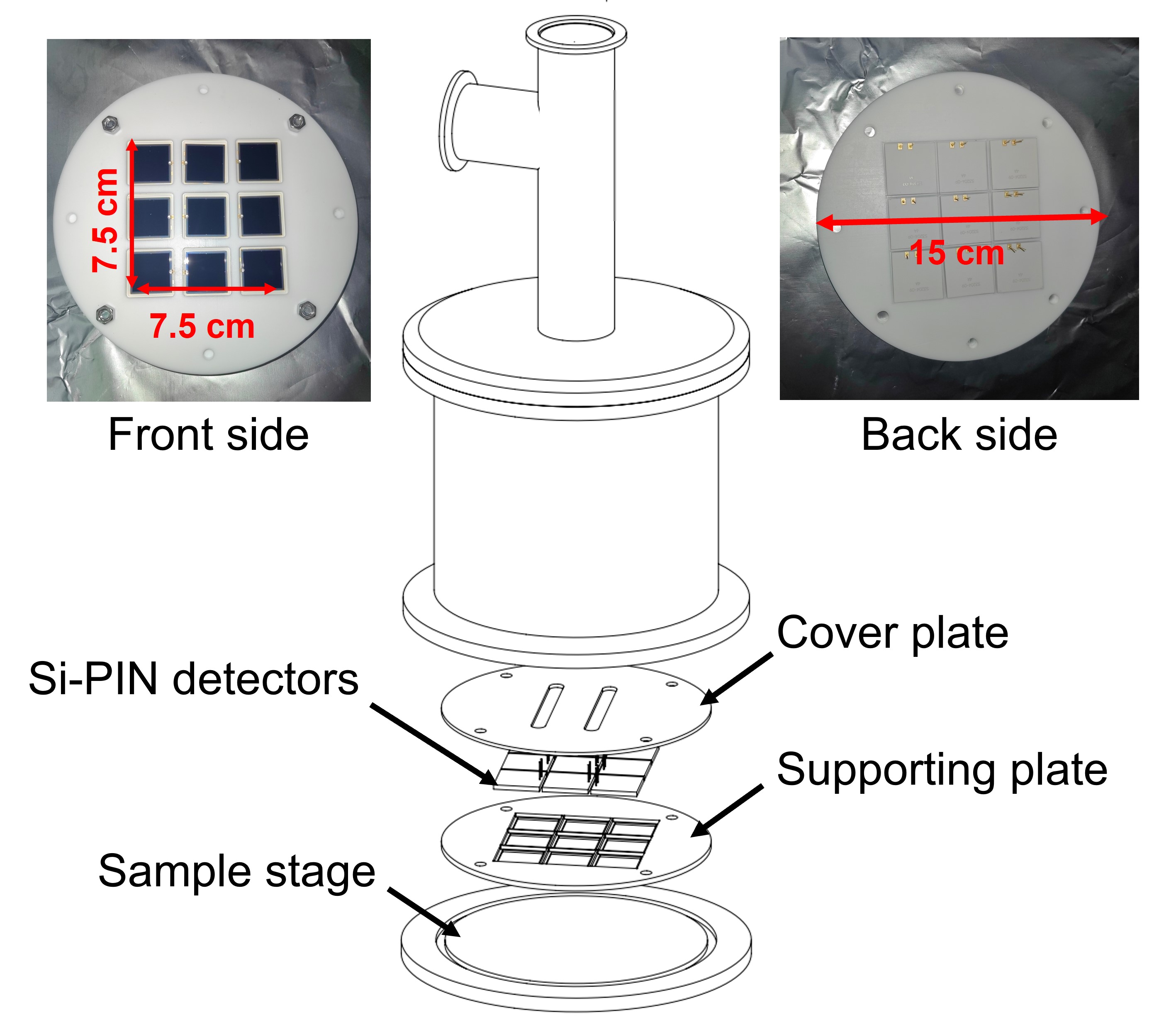}
		\caption{A detailed view of the sample chamber.}
		\label{1}
	\end{figure}

Fig.~\ref{1} shows a detailed view of the sample chamber. The chamber is constructed from a KF200 two-way flange with a height of approximately 20~cm, serving as the core enclosure for the sample and the Si-PIN detector. Its bottom is sealed with a KF200 blind flange, which also functions as the sample stage. The top is fitted with a custom adapter that converts a single KF200 port into two KF40 ports. Of these, one is connected to the vacuum pump via a KF40-to-1/4‑inch VCR adapter and a VCR valve. The other port is equipped with a KF40 SMA vacuum feedthrough and then connects to the electronics box through an SMA panel feedthrough on the glove box. One coaxial SMA cable is used to supply bias voltage to the Si-PIN detector and to carry the signal output.

Inside the chamber, a 3 $\times$ 3 array of nine Si-PIN detectors is installed for $\alpha$-particle detection. Each Si-PIN detector has an active area of 1.8 $\times$ 1.8~cm$^2$. The array is arranged in a square pattern with a side length of 7.5~cm, matching the maximum allowable sample size of 7.5 $\times$ 7.5~cm$^2$. The total active area of the detector array is 29.16~cm$^2$. All nine Si-PIN detectors are mounted on a PTFE support plate with their sensitive surfaces facing the sample stage. A PTFE cover plate is placed on the opposite side of the detectors, and the two plates are fastened together with screws to prevent any displacement during assembly or handling. Both the support plate and the cover plate have an outer diameter of 15~cm. Electrically, all nine Si-PIN detectors share a common bias supply, and their signal outputs are connected in parallel.

  \subsubsection{Glove Box}

The glove box, fabricated from acrylic, comprises a main chamber and an antechamber. The main chamber houses the sample chamber and provides a continuously stable, low-background, ultra-clean environment. The antechamber serves to facilitate safe sample transfer while preserving the internal atmosphere of the glove box. An external door on the left side of the main chamber allows for initial placement of the sample chamber, while an internal door on the right connects to the antechamber.

The main chamber is equipped with gas inlet and outlet ports, both fitted with quick-connect valves. The inlet is permanently connected to the vapor phase port of a liquid nitrogen tank, with boil-off nitrogen continuously supplied at a flow rate of 1~L/min to maintain a clean, low-radon nitrogen atmosphere inside. The outlet vents to the atmosphere through a check valve. The antechamber is also fitted with its own gas inlet and outlet. The inlet is connected to the main chamber’s gas supply line via a 10~mm quick-connect tee. After sample loading, the antechamber is purged with boil-off nitrogen to remove air, thereby ensuring the purity of the main chamber environment.

Two panel feedthroughs are installed on the main chamber: a 1/4-inch VCR panel feedthrough and an SMA panel feedthrough. The inner side of the VCR feedthrough is connected to the sample chamber via a stainless steel bellow, while the outer side is linked to the vacuum pump through another bellow. The inner side of the SMA feedthrough is wired to the Si-PIN detector, and the outer side is connected to the electronics box for bias supply and signal readout.

\subsubsection{Electronics}

\begin{figure}[htb]
		\centering
		\includegraphics[width=10cm]{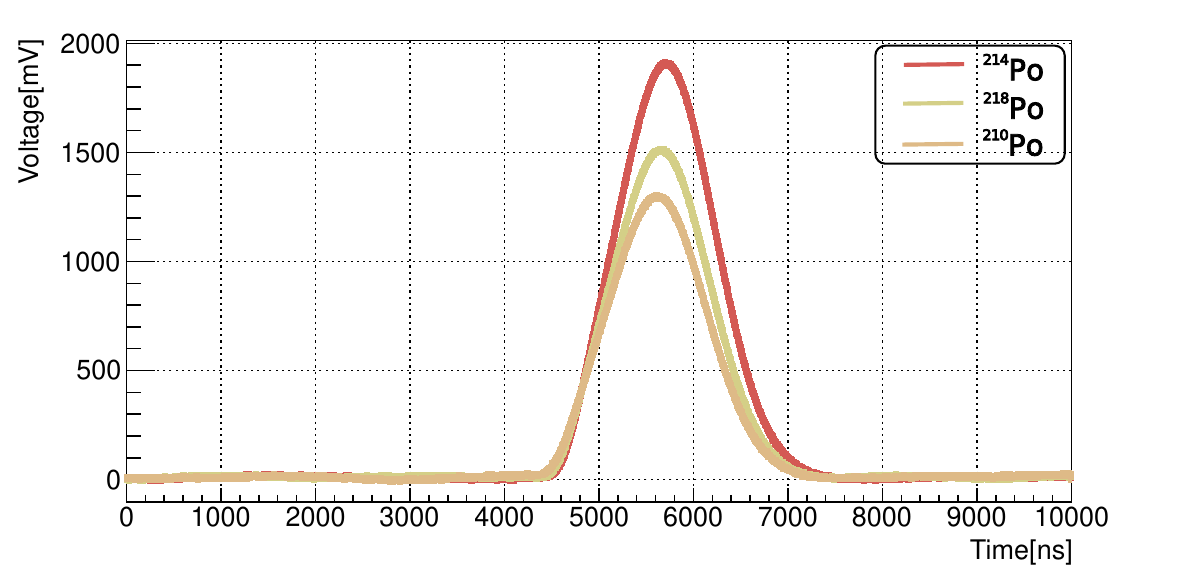}
		\caption{Sample waveforms recorded by the oscilloscope.}
		\label{waveform}
	\end{figure}

The SiPIN detectors share a single coaxial cable for both power supply and signal readout. The +35~V operating bias required by the SiPIN detectors is supplied by a DH1765 DC power supply~\cite{citekey3}, which offers a regulation precision of 1~mV. The SiPIN's signals pass through a 142A preamplifier (ORTEC)~\cite{citekey1} and a 671 main amplifier (ORTEC)~\cite{citekey2} before being fed into the oscilloscope (HDO6054, LeCroy)~\cite{citekey4}. The oscilloscope operates at a sampling rate of 250~MS/s with a sampling window length of 10~$\mu$s. Example pulses recorded by the oscilloscope are shown in Fig.~\ref{waveform}.

\subsubsection{Other Components}

In addition to the components mentioned above, the system incorporates a vacuum pump (ACP40, Pfeiffer) to evacuate the sample chamber, a 200~L liquid nitrogen tank to supply boil-off nitrogen for the measurement system, a KF40 valve and a 1/4-inch VCR valve to regulate gas flow, and a VCR pressure gauge to monitor the vacuum level inside the sample chamber.

	
	\subsection{System Operation}
To measure with the system, the sequential steps are outlined below:

1. System purge: Connect the nitrogen supply line and initiate a continuous purge of the main glove box using vaporized nitrogen. During operation, the inlet and outlet ports of the main chamber must remain open (except during liquid nitrogen tank replacement) to maintain a clean, low-radon internal environment.

2.  Initialize electronics: Switch on the preamplifier, main amplifier, and oscilloscope in sequence. Set the oscilloscope to a sampling rate of 250~MS/s, a trigger threshold of 350~mV, and a pulse acquisition time window of 10~$\mu$s.

3. Sample Loading and purging: For sample measurement, open the external door of the transfer chamber, place the sample inside, and then open its inlet and outlet ports. Purge the transfer chamber with boil-off nitrogen gas for at least 5 minutes.

4. Sample transfer: Following the purge, close all ports of the transfer chamber to seal it. Then, open the internal door connecting the main and transfer chambers, and transfer the sample into the main chamber.

5. Sample and detector assembly: Open the KF200 flange at the bottom of the sample chamber and place the sample onto the internal stage. Then, reinstall the KF200 flange and ensure a vacuum-tight seal using the designated O-ring and clamp.

6. Chamber evacuation: Once the sample chamber is sealed, open the KF40 valve and the 1/4 VCR valve sequentially. Start the vacuum pump to evacuate the sample chamber until the pressure gauge reads -101~kPa.

7. Data acquisition: Turn on the DC power supply and apply a +35~V bias voltage to the Si-PIN detector array and then save the data for subsequent analysis.

	\section{Detector calibration}\label{sec.3}

	\subsection{Radon Chamber}

	\begin{figure}[htb]
		\centering
		\includegraphics[width=11cm]{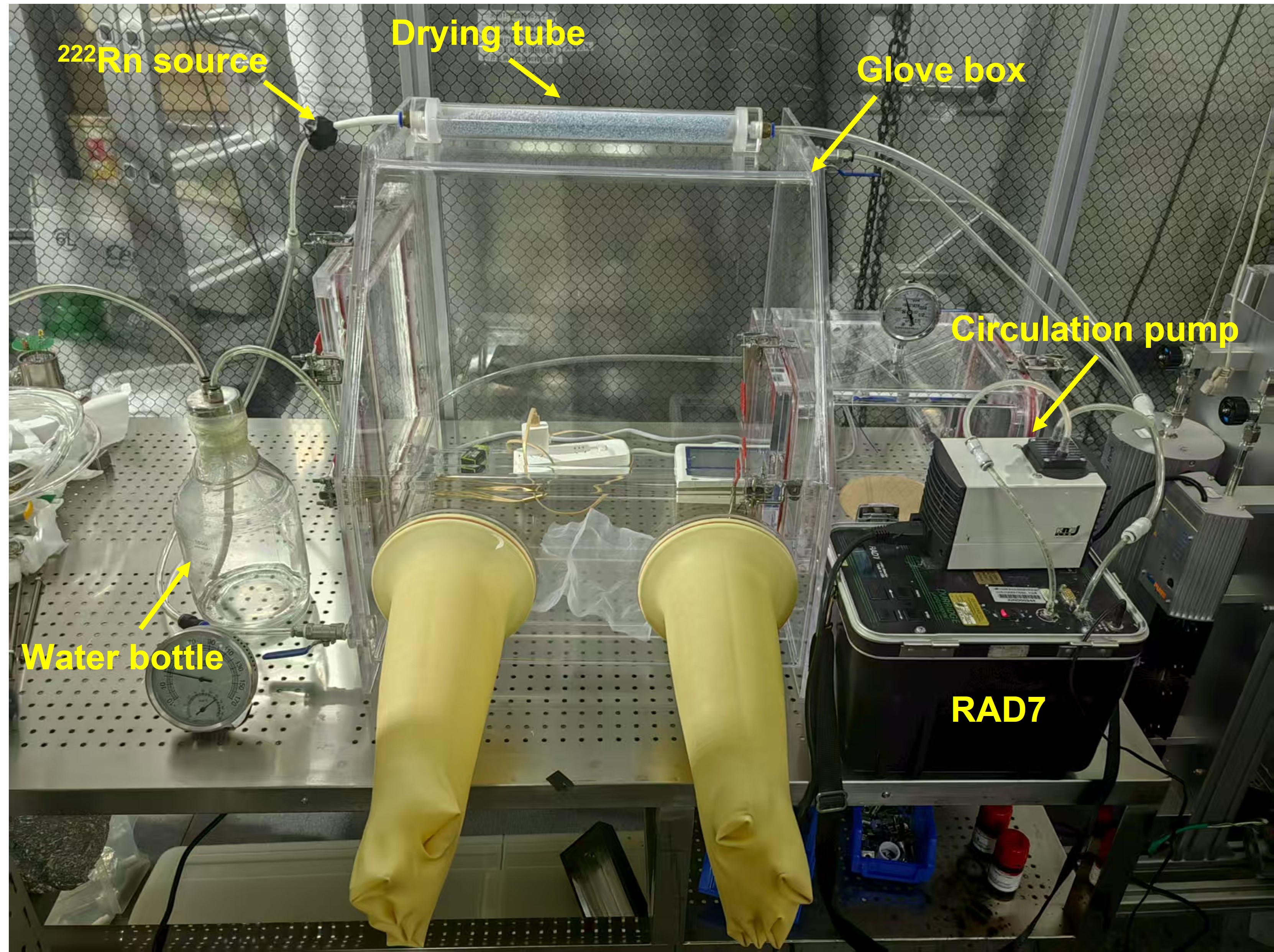}
		\caption{A photograph of the radon chamber.}
		\label{radon chamber}
	\end{figure}	

A high-concentration radon chamber was designed and constructed for the calibration of surface $\alpha$-activity measurement detector. A photograph of the chamber is presented in Fig.~\ref{radon chamber}. The system comprises a gas-flow radon source (activity $\sim$5000~Bq, fabricated by the University of South China)~\cite{calibration,calibration1,calibration2,calibration3}, a $\sim$500~L glove box serving as the main chamber, a circulation pump (N022AT, KNF), a drying tube (30~mm diameter $\times$ 400~mm length) containing $\sim$150~g of calcium sulfate desiccant, a water bottle, and a RAD 7 radon detector\cite{radon detector}.

 	\begin{figure}[htb]
		\centering
		\includegraphics[width=12cm]{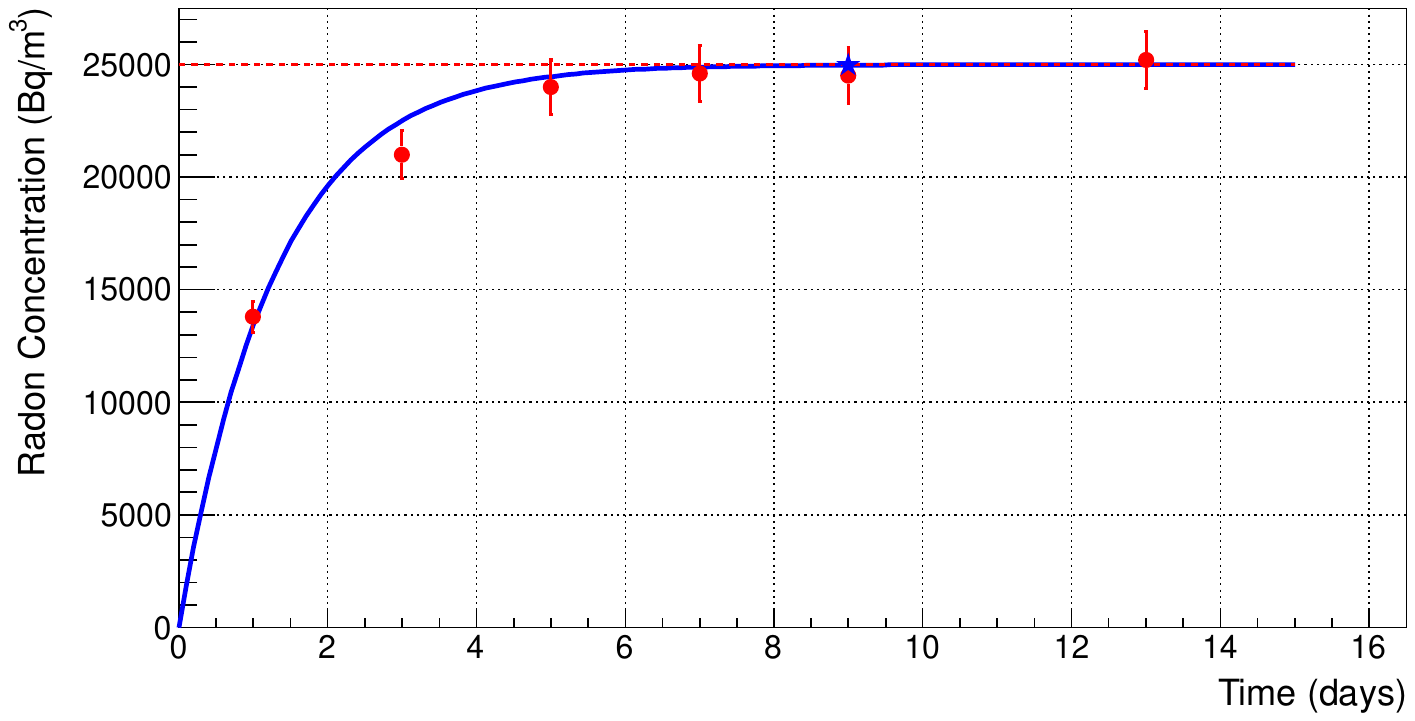}
		\caption{The change of radon concentrations inside the radon chamber. The red dots with error bars are measured by the RAD7 radon detector, and the blue line is an evolutionary curve added for ease of observation.}
		\label{radon}
	\end{figure}

Before initiating the gas cycle, the glove box is purged with boil-off nitrogen for at least five volumes to displace the air inside the glove box. The circulation pump then continuously circulated gas in the chamber through the radon source, gradually increasing the radon concentration. The radon concentration was monitored in real time using the RAD 7 radon detector. As shown in Fig.~\ref{radon}, the concentration reached an equilibrium value of $\sim$25000~Bq/m$^3$ after 9 days. This stable radon atmosphere provided a uniform radioactive environment for all subsequent deposition experiments.

For detector calibration, PMMA plates with a diameter of 15~cm were chosen as the standard sample material. PMMA is widely used in low-background particle physics experiments—such as JUNO ~\cite{juno} and SNO+~\cite{maneira2013sno+}—where it serves as a key structural component in central detectors. In this work, PMMA plates were placed inside the radon chamber to allow for the deposition of radon decay progeny, notably $\alpha$-emitting isotopes like $^{214}$Po and $^{210}$Po, onto their surfaces. This process produced well-defined $\alpha$ sources with known activity levels. These calibrated sources were then used to systematically perform the surface $\alpha$-activity detector calibration and to assess its key performance characteristics, including energy response and energy resolution.

 	\subsection{Energy Response}

	\begin{figure}[htb]
		\centering
		\includegraphics[width=12cm]{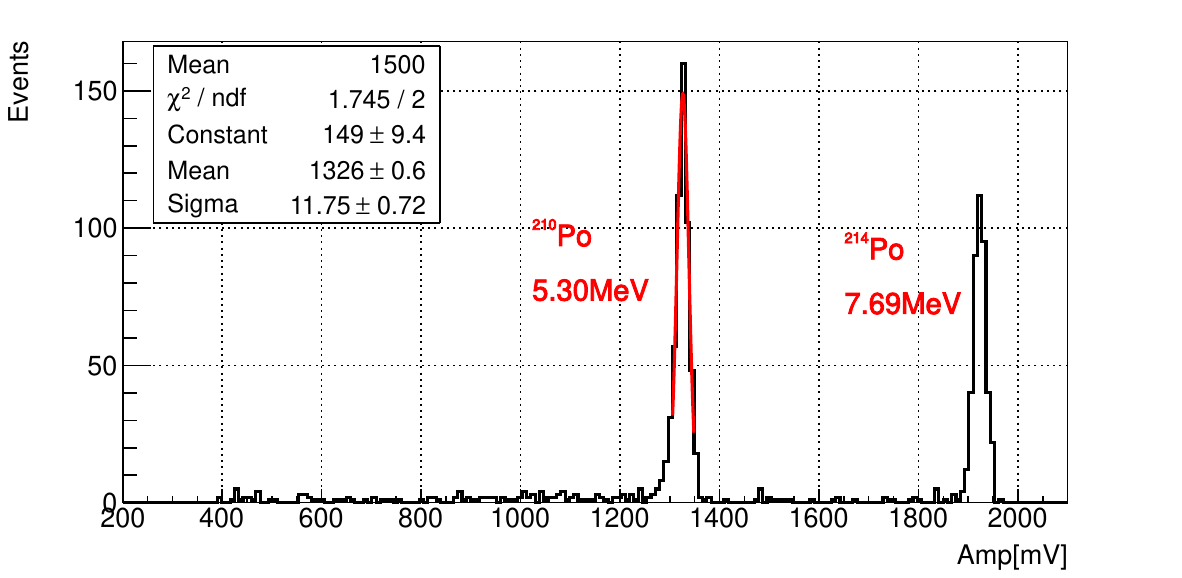}
		\caption{The amplitude spectrum measured by the surface alpha activity measurement system and the peaks caused by $^{210}$Po and $^{214}$Po decay $\alpha$s are indicated. Given that this system was primarily developed to measure the activity of $^{210}$Po, the $\alpha$ peak generated by $^{210}$Po was fitted with a Gaussian function, and the fitting parameters as shown.}
		\label{Amp}
	\end{figure}

The energy calibration of the surface alpha measurement system was performed using a PMMA plate that had been exposed in the radon chamber for around two weeks. The measured amplitude spectrum is shown in Fig.~\ref{Amp}. Given the short half-life of $^{218}$Po (3.1~min), the detected $\alpha$ particles are predominantly from the decays of $^{214}$Po and $^{210}$Po. As the primary goal of this apparatus is to quantify $^{210}$Po surface activity, the corresponding $\alpha$ peak was fitted with a Gaussian function. The fit yielded a signal amplitude of 1325.70 $\pm$ 0.61~mV for the 5.30~MeV $\alpha$ particles from $^{210}$Po. Furthermore, the energy resolution of the detector at 5.30~MeV was determined to be 2.09 \% based on the fitting results.

   \subsection{Sensitivity estimation}

	The sensitivity of measuring the $^{210}$Po activity on sample surface at 95\% confidence level can be estimated according to Eq.~\ref{sensitivity}~\cite{31,30-2}:
    \begin{equation}\label{sensitivity}
    L = \frac{1.645 \times \sigma_{\mathrm{BG}}}{C_F \times S \times 10^6}
    \end{equation}
where L is the measurement sensitivity in the unit of $\mu$Bq/cm$^2$, $\sigma_{BG}$ is the statistical uncertainty of the system's background in the unit of Counts Per Day (CPD), C$_F$ is the calibration factor of the detector, S is the area of the Si-PIN sensitive surface, which is 29.16~cm$^2$. 

The detector sensitivity was calculated using the effective area of the Si-PIN detector rather than the sample area. Given that the detector-sample distance is less than 1 cm and both are placed in a vacuum, the $\alpha$ detection efficiency was assumed to be 100\%, corresponding to a Calibration Factor (C$_F$) of 8.64 $\times$ 10$^{-2}$~CPD/$\mu$Bq.

    \begin{figure}[htb]
	\centering
	\includegraphics[width=12cm]{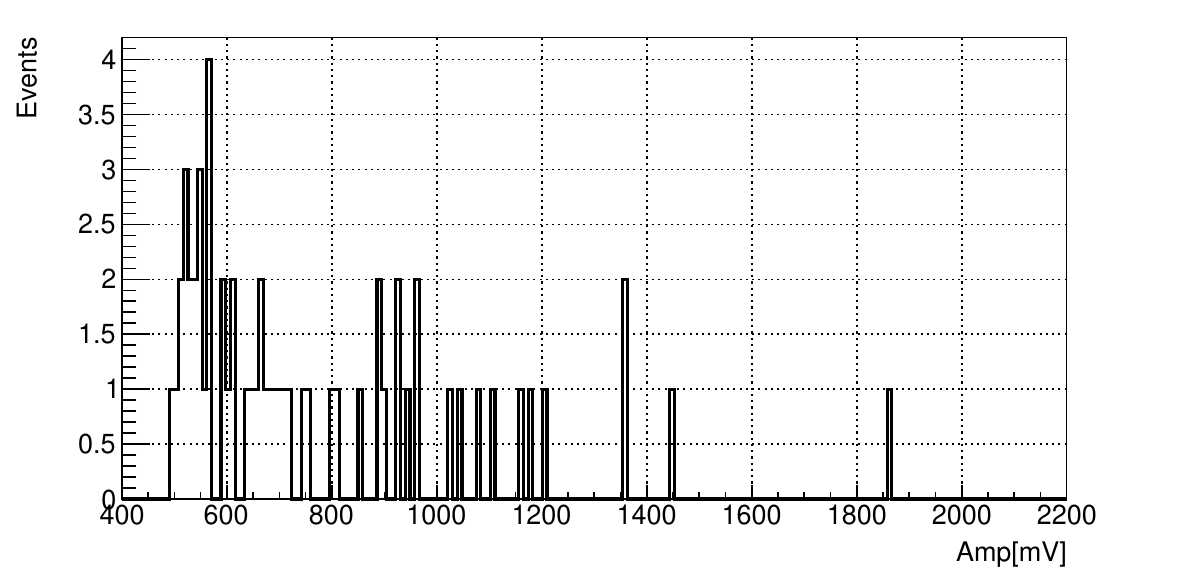}
	\caption{The amplitude spectrum of the background. The amplitude range of interest is 1297~mV to 1351~mV.}
	\label{background}
	\end{figure}

\begin{figure}[htb]
	\centering
	\includegraphics[width=12cm]{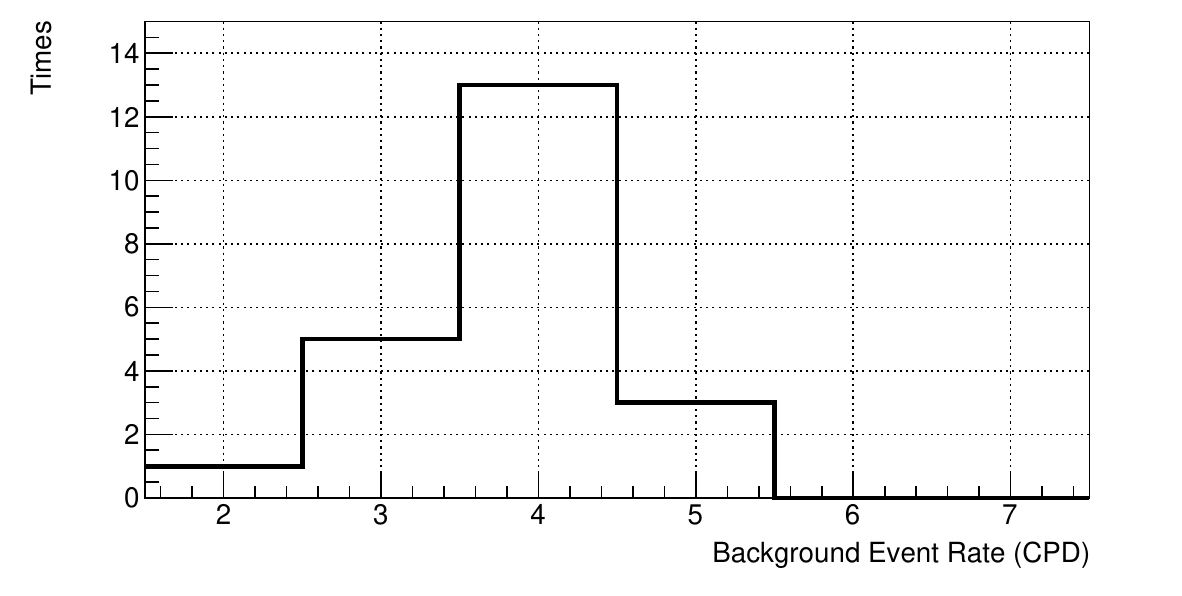}
	\caption{Distribution of twenty-one single-day background event rates.}
	\label{bendi}
	\end{figure}

The system background was evaluated by measuring the count rate without any sample. An example amplitude spectrum obtained under these conditions is shown in Fig.~\ref{background}. The region of interest for $^{210}$Po was defined as the mean $\pm$3$\sigma$ interval, corresponding to amplitudes between 1297~mV and 1351~mV. Each measurement was conducted over a 24-hour acquisition period. Twenty-two repeated measurements yielded the background count rate distribution shown in Fig.~\ref{bendi}, with an average value of 3.82 $\pm$ 1.95~CPD, where the uncertainty of 1.95 represents the statistical error derived from the Poisson distribution.

Using the parameters mentioned above and calculated with Eq.~\ref{sensitivity}, a one-day measurement sensitivity of 1.27 $\mu$Bq/cm$^2$ for the surface $^{210}$Po activity is derived.

\section{Measurement results}\label{sec.4}

Under controlled experimental conditions, this study investigated the effects of exposure time, surface electrostatic potential, and surrounding humidity on the deposition behavior of radon progeny onto PMMA surfaces. Given that $^{214}$Po is preceded by $^{214}$Pb with a half-life of 26.8~min and $^{214}$Bi with a half-life of 19.9~min, the $\alpha$ event rate from its decay is selected as the reference quantity for evaluating the attachment of radon progeny on the PMMA surface.

As illustrated in Fig.~\ref{radon chamber}, a water bottle is connected to the gas-circulating circuit to regulate the humidity inside the radon chamber. A drying tube is installed between the radon chamber and the RAD 7 detector, as the measurement accuracy of the RAD 7 is strongly influenced by the humidity of the sampled gas. It is therefore essential to maintain the relative humidity of the gas entering the RAD 7 below 10\%.

The electrostatic potential on the acrylic surface was generated by rubbing the PMMA plate with a polyester microfiber dust‑free cloth, which imparts a negative charge to the surface. The resulting surface potential was measured using an electrostatic tester (JH‑TEST, Dongguan Zejing Electronic Technology Co., Ltd.).

Before each experiment, the PMMA sample surface was cleaned in an ultrasonic cleaner and dried by blowing boil-off nitrogen gas over it. The sample was then placed into the radon chamber and exposed to $\sim$25000~Bq/m$^3$ high-radon environment. After exposure, it was transferred to the surface $\alpha$‑activity measurement system for $^{214}$Po event rate counting. Following background correction, the $^{214}$Po counting rate was calculated using Eq.~\ref{AAA}:
   \begin{equation}\label{AAA}
    A = \frac{ N - N_{\mathrm{BG}}}{t}
    \end{equation}
where A is the $^{214}$Po counting rate in the unit of Counts Per Hour (CPH), N is the events of detected $\alpha$ particles from $^{214}$Po decay in the unit of counts, N$_{BG}$ is the background event rate at the energy range of $^{214}$Po $\alpha$ in the unit of counts, and t is the data-taking time in the unit of hour. Each measurement lasted 3 hours, covering several half‑lives of $^{214}$Pb and $^{214}$Bi. This ensures that the majority of $^{214}$Po atoms are formed and decay within the measurement period, thereby guaranteeing that the recorded counts fully represent the total deposited radon progeny.

	\subsection{Exposure time}

The deposition of radon progeny is a dynamic process in which the deposition activity does not increase monotonically, but is governed by the interplay of deposition, radioactive decay, and desorption.

\begin{figure}[htb]
\centering
\includegraphics[width=12cm]{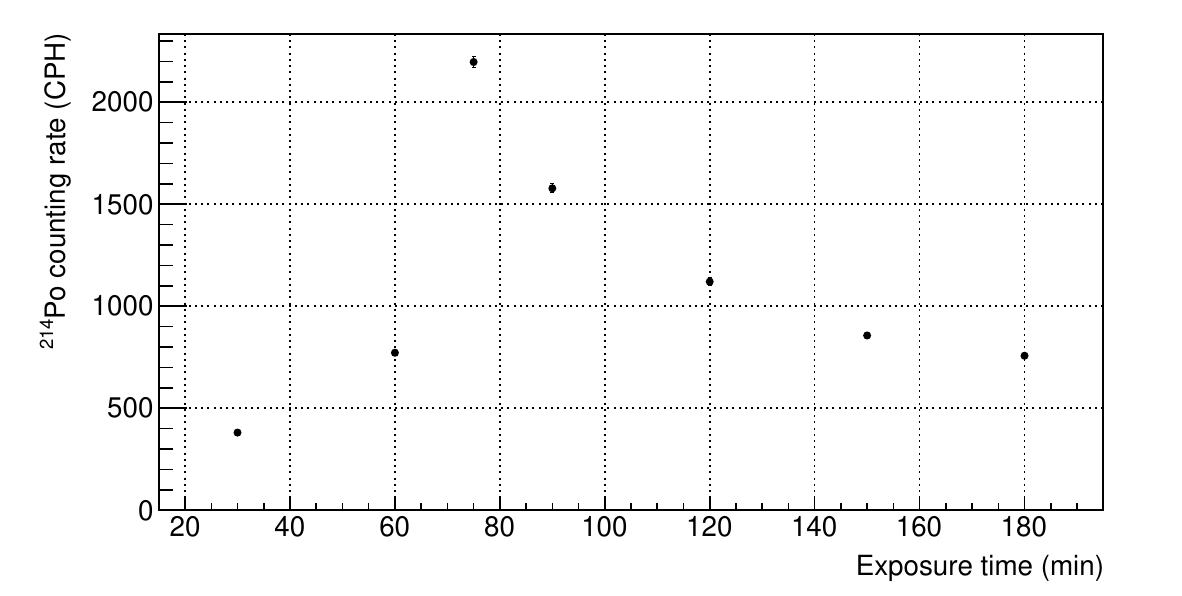}
\caption{Deposited $^{214}$Po counting rate as a function of exposure time. The errors in the Y-axis are statistical only, and the error for exposure time is negligible.}
\label{time}
\end{figure}

As shown in Fig.~\ref{time}, the measured $^{214}$Po event rate varies non-monotonically with exposure time, exhibiting a distinct peak at ~75 minutes. The initial rise in event rate is attributed to a transient equilibrium established between the continuous deposition of radon progeny and the successive decays within the chain. Beyond the peak, the deposition activity gradually declines. This decrease is primarily influenced by two factors: (1) the production rate of $^{214}$Po, which is controlled by its parent nuclide $^{214}$Bi, and (2) the increasingly significant effect of $\alpha$-recoil-induced desorption ~\cite{fleischer1988alpha,recoil}. As more progeny atoms deposit on the surface, the recoil momentum from subsequent $\alpha$-decays causes a fraction of the already-adhered atoms to detach, resulting in a net reduction of the steady-state activity.

	\subsection{Surface electrostatic potential}

    \begin{figure}[H]
	\centering
	\includegraphics[width=12cm]{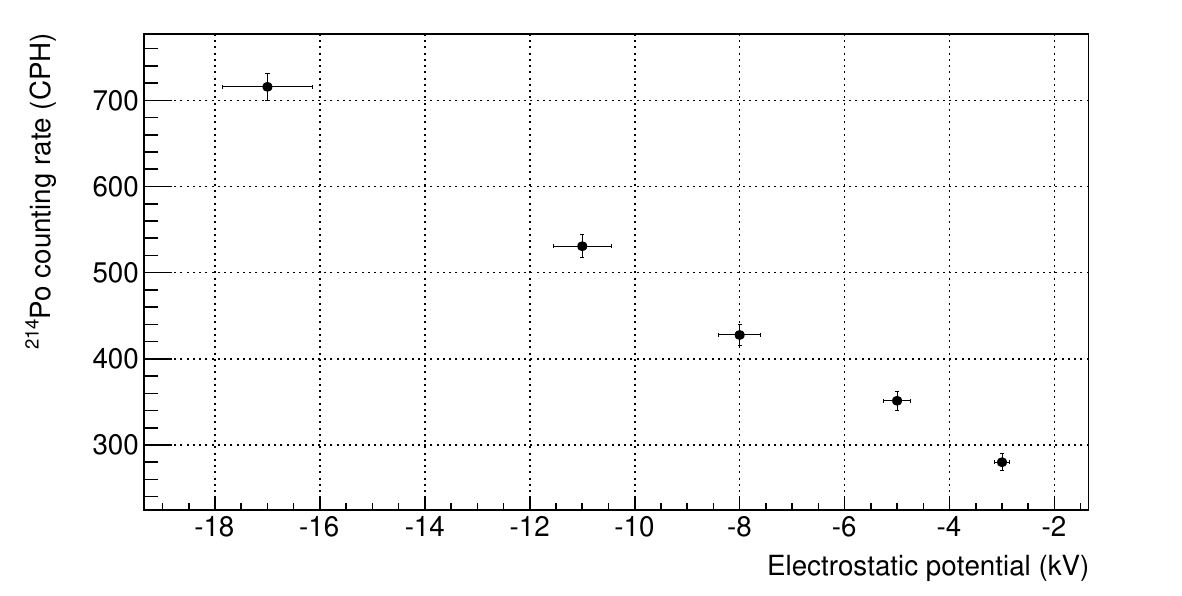}
	\caption{The relationship between the deposited $^{214}$Po counting rate and surface electrostatic voltage; the error on the Y-axis is the statistical error, while the errors on the X-axis are the uncertainties of the electrostatic meter ( which is $\pm$5\% according to its instruction).}
	\label{charge}
	\end{figure}

	Since more than 90\% of radon progeny are positively charged~\cite{SuperK1999,10.1093/ptep/ptv018,SNO}, and electrostatic fields directly influence the migration and attachment of charged ions\cite{ele1,ele2,ele3}, the surface deposition activity of radon progeny is highly sensitive to the surface electrostatic potential~\cite{ele}.

    As shown in Fig.~\ref{charge}, the measured $^{214}$Po event rate on the surface of an acrylic panel, which was exposed to the radon chamber for a standard 5-minute period, exhibits a clear dependence on electrostatic potential. Within the investigated voltage range (approximately -3~kV to -17~kV), the deposition activity increases continuously with the magnitude of the negative potential. The $^{214}$Po event rate rises from about 280~CPH at -3~kV to around 715~CPH at -17~kV. This trend indicates that, within the studied voltage range, negative electric fields produce electrostatic attraction of positively charged radon progeny, thereby promoting their deposition on the material surface. The stronger the field, the higher the collection efficiency for positive ions, resulting in a monotonic increase in deposition activity.

\subsection{Humidity}
The surface deposition activity of radon progeny is also highly sensitive to ambient humidity, primarily because humidity affects both the distribution of surface charges on insulating materials such as PMMA and the charge‑neutralization of the positively charged progeny.

	\begin{figure}[htb]
		\centering
		\includegraphics[width=12cm]{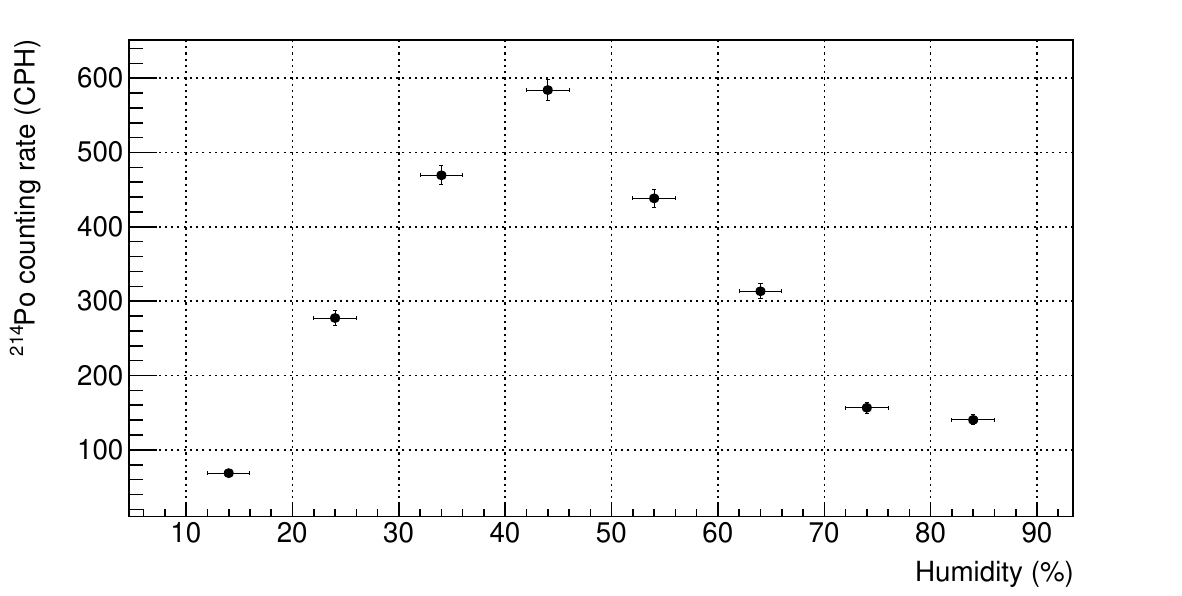}
		\caption{The relationship between deposited $^{214}$Po counting rate and humidity; the errors on the Y-axis are statistical only, while the error on the X-axis are the uncertainty of the hygrometer (which is $\pm 2\%$ according to its instruction).}
		\label{humidity}
	\end{figure} 

As shown in Fig.~\ref{humidity}, the deposited $^{214}$Po event rate after a standard 30‑minute exposure shows a clear dependence on the humidity inside the radon chamber.

Under low‑humidity conditions, the measured count rates are significantly suppressed. This is attributed to the non‑uniform distribution of surface charges on the acrylic sheet~\cite{ele4,elehum}. In dry air, charges remain localized mainly at discrete friction points, limiting the effective adsorption of radon progeny to confined regions. As humidity increases, moisture facilitates a more uniform redistribution of surface charges across the acrylic surface. This expands the effective adsorption area, allowing deposition activity to increase progressively and reach a maximum at around 44\% relative humidity. Beyond this optimum, further increases in humidity lead to a decline in measured activity. At elevated humidity levels, water molecules—acting as polarizable or charge‑carrying species—can neutralize the positively charged radon progeny in the gas phase~\cite{Microbubble}. Once neutralized, these particles are no longer attracted electrostatically to the negatively charged acrylic surface, resulting in reduced collection efficiency.

	\section{Summary}\label{sec.5}

In low-background experiments such as direct dark matter detection, the deposition of radon progeny on detector surfaces constitutes a significant background source. In particular, the long-lived $^{210}$Pb and its decay product $^{210}$Po can produce persistent background interference through $\alpha$ decay and ($\alpha$,n) reactions. 

To accurately measure the $\alpha$ activity on material surfaces, a high-sensitivity surface $\alpha$-activity measurement system has been developed. The system employs an array of nine Si-PIN detectors to perform $\alpha$ spectroscopy on sample surfaces under vacuum conditions. The detector was calibrated in energy and sensitivity, achieving a one-day measurement sensitivity of 1.27~$\mu$Bq/cm$^2$ for $^{210}$Po surface activity at the 95\% confidence level. Using a self-constructed high-concentration radon chamber, the deposition behavior of radon progeny on PMMA surfaces was systematically investigated under different conditions. Experimental results show that the deposition activity varies non-monotonically with exposure time, peaking at approximately 75 minutes; increasing negative electrostatic potential on the surface significantly enhances the deposition of radon progeny; and ambient humidity strongly affects deposition efficiency, with maximum activity observed at around 44\% relative humidity. 

This study provides important experimental insights for the control and removal of surface radon contamination in low-background experiments.

	\section{Acknowledgement}
	This study is partly supported by the National Natural Science Foundation of China (Grant No. 12275289) and the Youth Innovation Promotion Association of the Chinese Academy of Sciences (2023015).
	
	\section{Conflict of interest}

On behalf of all authors, the corresponding author states that there is no conflict of interest.

\end{document}